\documentclass[preprint]{iopart}

\usepackage{epsfig}

\begin{document}

\title{Evidence for compact cooperatively rearranging regions  in
  a supercooled liquid}

\author{M. Elenius$^1$ and M. Dzugutov$^2$}

\address{$^1$ Dept. for Numerical Analysis and Computer Science\\ 
Royal Institute of Technology, S--100 44 Stockholm, Sweden}
\address{$^2$ Dept. for Materials Science and Engineering\\ 
Royal Institute of Technology, S--100 44 Stockholm, Sweden}
\ead{mik@pdc.kth.se}

\begin{abstract}

  We examine structural relaxation in a supercooled glass-forming
  liquid simulated by NVE molecular dynamics. Time correlations of the
  total kinetic energy fluctuations are used as a comprehensive
  measure of the system's approach to the ergodic equilibrium. We find
  that, under cooling, the total structural relaxation becomes delayed
  as compared with the decay of the component of the intermediate
  scattering function corresponding to the main peak of the structure
  factor. This observation can be explained by collective movements of
  particles preserving many-body structural correlations within compact 3D
  cooperatively rearranging regions.

\end{abstract}

\submitto{\JPCM}

\maketitle

Understanding the microscopic mechanisms of the super-Arrhenius
slowing down in fragile glass-forming liquids \cite{Angell91} remains
a major problem in condensed-matter physics
\cite{Ediger96,Debenedetti}. It is understood that below the
temperature $T_A$ that marks the crossover to super-Arrhenius
behaviour liquid dynamics is controlled by the topography of the
potential energy-landscape (PEL) \cite{Goldstein}. The PEL is known to
be divided into metabasins - areas of PEL confining sets of mutually
well-connected local energy minima \cite{Stillinger}. Inter-metabasin
transitions, mediating the (primary) $\alpha-$relaxation, involve
uncorrelated movements of large groups of particles \cite{Kob}. On a
shorter time scale, confinement to a metabasin imposes structural
constraints upon the atomic motions, reducing the number of accessible
degrees of freedom. This renders intra-metabasin dynamics highly
collective, in accordance with the Adam-Gibbs concept of independent
cooperatively rearranging regions (CRR) \cite{Adam}.

The geometry, and dynamics of CRR remain elusive. By
definition\cite{Adam}, a CRR is a minimum-size region of structure
possessing a relaxational degree of freedom independent of the region's
environment. It was suggested \cite{Buldyrev} that this degree of
freedom can be associated with a coherent linear translation of a
string-like cluster of particles within slowly-changing structural
environment\cite{Donati}. On the other hand, the random first-order
transition theory of glasses \cite{Wolynes} predicts a transition, at
some stage of the liquid's cooling, from a string-like shape of CRR to
a compact one. The concept of a compact CRR presumes that there exists
a relaxational degree of freedom that makes it possible for a 3D
configuration of particles confined to that CRR to perform a
cooperative movement that would preserve some of the many-body
structural correlations constraining the configuration.

A distinctive aspect of the relaxation dynamics associated with
compact CRR is that the described cooperative 3D movements of
particles are expected to result in decorrelation of the density
fluctuations while preserving slower-decaying many-body structural
correlations. The relaxation of the density fluctuations is described
by the intermediate scattering function $F(Q,t)$\cite{Hansen}. Its
slowest-decaying component (within the relevant range of $Q$)
corresponds to $Q_m$ - the position of the main maximum of the
structure factor $S(Q)$. In the case of cooperative 3D particle
movements within compact CRR, $F(Q_m,t)$ is expected to decay well
before the total liquid's structural relaxation has been accomplished.
This can be understood using the following simple model. Consider
slowly relaxing structural domains (clusters) immersed in a faster
relaxing liquid. Let the clusters perform uncorrelated rotational and
translational movements. It is easy to see that $F(Q_m,t)$ for such a
model would decay well before the clusters dissipate as structural
entities.

Detecting the described effect requires the use of a comprehensive
measure of the structural relaxation that would be independent of the
dissipation of density fluctuations. In this paper, we report a
molecular dynamics (MD) simulation of a fragile glass-forming liquid
where the structural relaxation is measured by the time correlations
of the fluctuations of the system's total kinetic energy. We find
that, as the liquid is cooled sufficiently close to the mode-coupling
theory (MCT) \cite{gotze} critical temperature $T_c$, its total
structural relaxation becomes delayed as compared with the respective
decay of $F(Q_m,t)$. This anomaly can be interpreted as an indication
of the existence of slowly dissipating many-body correlations
constraining the movements of cooperatively rearranging 3D groups of
particles. These groups can be identified with compact CRR.

The MD simulation we report here explores a fragile simple
one-component glass-forming liquid demonstrating a pronounced tendency
for icosahedral clustering \cite{Z2} (named Z2 in that reference),
with the estimated $T_c=0.65$, fragility index $B=4.5$, and $T_A=1.1$.
For each temperature, an equilibrium simulation in an $NVE$ ensemble
was performed using a system of 16384 particles. A system of 128000
particles was also tried, and no size-dependent effects have been
observed.

\begin{figure} 
  \includegraphics[width=7.9cm]{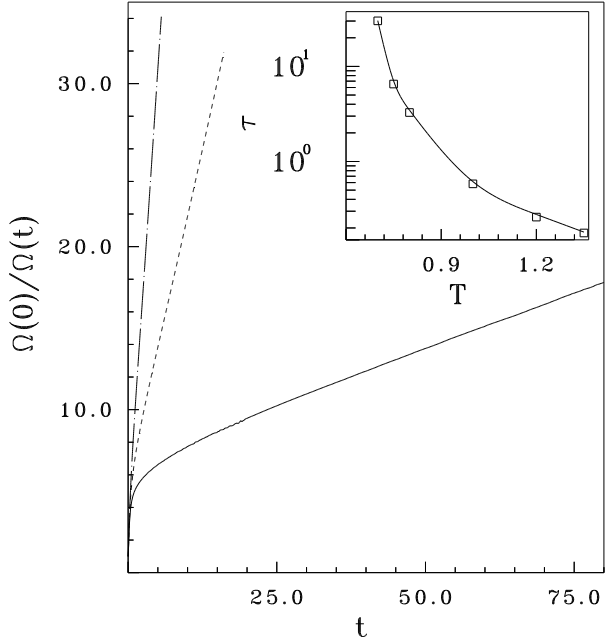}
  \caption{ Asymptotic behaviour of $\Omega(t)$ as predicted by
    \ref{omegaeq}. Solid line, $T=0.75$; dashed line, $T=1$; chain-dotted
    line, $T=1.35$. Inset: solid line, $\tau$ as evaluated from the
    asymptotic slope of $\Omega(0)/\Omega(t)$, defined in \ref{omegaeq}; boxes,
    $\tau$ calculated by integrating $c(t)$, defined in \ref{taueq}}
    \label{fig1}
\end{figure}

Consider a system of $N$ particles in an $NVE$ ensemble.  Using the
expression for the isometric heat capacity \cite{Lebowitz, Haile}, its
entropy $s$ can be linked to the variance of the kinetic energy $k$
(both per particle):

\begin{equation} 
  c_v^{-1} =  \frac{1}{T}  \left ( \frac{ \partial s} { \partial T} \right
  )_V^{-1} = \frac{2}{3} -  N \frac{ \langle k^2 \rangle
  - \langle k \rangle^2 } {  \langle k \rangle^2 }
\end{equation}
The respective expression for the entropy explored by the system
within the time interval $t$, $s(t)$, \cite{Ma}, will be:

\begin{equation} 
  \frac{1}{T}  \left ( \frac{ \partial s(t)} { \partial T} \right
  )_V^{-1} = \frac{2}{3} -  N \frac{ \langle \langle k^2 \rangle_t
  - \langle k \rangle_t^2 \rangle} { \langle k \rangle^2}
\end{equation}
where $ \langle ~ \rangle_t$ denotes averaging over the time
interval $t$.  The system's approach to the ergodic equilibrium can be
comprehensively assessed as the difference between the two quantities
\cite{Palmer}:

\begin{equation} 
 \Omega (t) =  \frac{1}{T}  \left [ \left ( \frac{ \partial s} { \partial T} \right
  )_V^{-1} - \left ( \frac{ \partial s(t)} { \partial T} \right
  )_V^{-1}  \right ] = N \frac{ \langle k\rangle^2 - \langle
  \langle k \rangle_t^2 \rangle } { \langle k \rangle^2}
\end{equation}
Note that $\Omega (t)$ is a formal analogue of an earlier suggested
measure of ergodic convergence \cite{Mountain} calculated from the
time-dependent variance of the {\it single} particle energy. Following
the arguments presented in that work, it is also possible to conclude
that the large-$t$ asymptotic behaviour of $\Omega (t)$ can
be described as
\begin{equation} 
 \Omega (t) /  \Omega (0)  =  \frac{ \langle k\rangle^2 - \langle\langle k
  \rangle_t^2\rangle} { \langle k \rangle^2 - \langle k^2 \rangle} 
 \longrightarrow \frac{ \tau }{ t},\: t \rightarrow \infty
 \label{omegaeq}
\end{equation}
where 
\begin{equation} 
 \tau = 2 \int_0^{\infty} dt \,c(t) / c(0)
 \label{taueq}
\end{equation}
and $c(t)$ is the time correlation function for the system's total
kinetic energy fluctuations:
\begin{equation} 
 c(t) = \langle \delta k(t) \delta k(0) \rangle
 \label{cteq}
\end{equation}
with $\delta k(t) = k(t) - \langle k \rangle$. This asymptotic
behaviour of $\Omega (t)$ is indeed observed in our MD simulations,
both above and below $T_A$, see Fig.~1. The inset of that figure
demonstrates the agreement between $\tau$ as evaluated from the
asymptotic slope of $\Omega (0)/\Omega (t)$ \ref{omegaeq},
and that obtained by direct integration of the correlation function
$c(t)$ \ref{taueq}.

\begin{figure} 
\includegraphics[width=7.9cm]{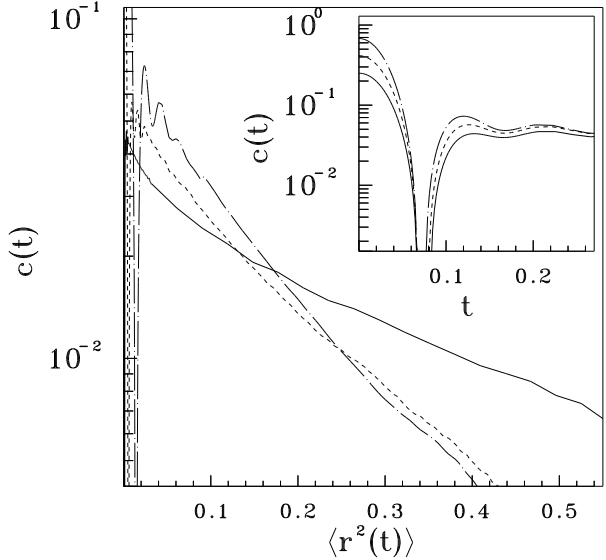}
\caption{Evolution of the correlation function $c(t)$, as defined
by \ref{cteq}, shown as a function of the mean-square
displacement. Solid line, $T=1$; dashed line, $T=0.75$;
chain-dotted line, $T=1.35$. Inset: the short-time behaviour of
the same curves. Note that in the inset they are shown as a
function of time.}
\label{fig2}
\end{figure}

\begin{figure} 
\includegraphics[width=7.9cm]{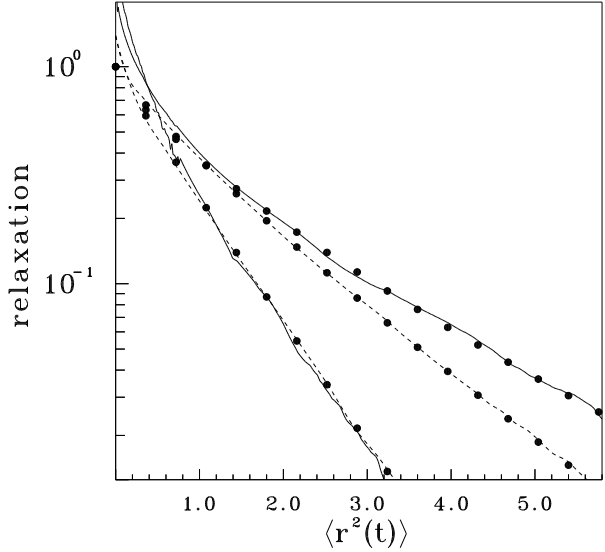}
\caption{ Structural relaxation in the temperature domain of
supercooled liquid dynamics. Solid lines, $c(t)$; dashed lines,
$F(Q_m,t)$; left and right curves, respectively, correspond to
$T=0.75$ and $T=0.7$. Dots: the fits of KWW stretched exponential
approximation $A \exp{[ - (t/ \tau)^\beta]}$. The values of the
fitting parameters are presented in Table 1. All sets of data
presented in the plot are scaled by the respective values of $A$.}
\label{fig3}
\end{figure}

Thus, the system's relaxation, as measured by $\Omega(t)$, is
ultimately controlled by the decay of $c(t)$. The general
structure of this correlation function is shown in Fig.~2. Its
short-time behaviour (shown in the inset) is dominated by
large-scale oscillations. The latter are evidently caused by particles
vibrations within the cages of their immediate neighbours, the
frequency being determined by the mean-square force \cite{Yip}
with  weak temperature dependence.

The described short-time behaviour of $c(t)$ obviously excludes the
possibility of using its normalised integral $\tau$, equation (5),
which determines the decay rate of $\Omega(t)$, as an unambiguous
measure of the liquid's relaxation rate. However, the initial
oscillatory regime of $c(t)$ is distinct in time, and decoupled from
the relaxational regime that follows it. The structural relaxation in
the asymptotic $\alpha$-regime can therefore be adequately measured by
the long-time decay of $c(t)$. We remark that, above $T_A$, this
longer-time relaxational evolution of $c(t)$, when presented as a
function of the mean-square displacement, Fig.~2, demonstrates the
standard liquid universality of the diffusion-relaxation relation.
This universality evidently breaks below $T_A$.

We now exploit $c(t)$ for a detailed analysis of the
$\alpha$-relaxation process in the supercooled dynamics regime. In
Fig.~3, the asymptotic behaviour of $c(t)$ is compared with that of
$F(Q_m,t)$, for two temperatures below $T_A$. For convenience, both
are plotted as a function of the mean-square displacement.  To
facilitate the comparative quantitative analysis, the correlation
functions presented in Fig.~3 have been fitted, within the relevant
domains of time, with the Kohlrausch-Williams-Watts (KWW) stretched
exponential function $A \exp{[ - (t/ \tau)^\beta]}$ which we use here
in the equivalent form $ A \exp{\{ - [ \langle r^2(t) \rangle /
6D\tau]^\beta\}}$, $D$ being the diffusion coefficient.  The
parameters' values for the fits are compiled in Table 1. Furthermore,
to make it possible to compare the two kinds of correlation functions
on equal footing, each function has been scaled by the respective
value of the KWW parameter $A$. Both $F(Q_m,t)$ and $c(t)$ appear to
be well described by the KWW approximation within the
$\alpha$-relaxation time domain. The latter follows the initial stage
of the relaxation process, confined to $\langle r^2(t)\rangle<1$, that
corresponds to conformational rearrangements of particles within the
cage of the first neighbours.
\begin{table} 
\vspace{0.5cm}
\begin{tabular}{|p{2.cm} |p{1.cm} |p{1.cm} |p{1.cm} |p{1.cm} |} \hline \hline

		&  $T$  & $A$	 & $6D\tau$ & $\beta$     \\ \hline
 $c(t)$		& 0.75	& 0.0393 & 0.71	  & $0.96$	\\ \hline
 $F(Q_m,t)$  	& 0.75	& 0.72	 & 0.71	  & $0.96$	\\ \hline
 $c(t)$		& 0.7	& 0.0557 & 1.02	  & $0.75$	\\ \hline
 $F(Q_m,t)$	& 0.7	& 0.72	 & 1.02	  & $0.865$	\\ \hline \hline 

\end{tabular}
\caption{ Parameters' values for the stretched-exponential KWW fits to
the relaxation functions shown in Fig. 3}
\label{tab3}
\end{table}

For $T=0.75$, the asymptotic behaviour of the two correlation
functions is indistinguishable, within the accuracy of the KWW fit
\cite{accuracy}. However, as the liquid is cooled to $T=0.7$, the
asymptotic decay of $c(t)$ becomes considerably delayed as compared
with the respective decay of $F(Q_m,t)$. We also note that the
observed delayed decay of $c(t)$ is accounted for by a measurable
reduction of the respective KWW stretching parameter $\beta$
(Table~1). The relaxation stretching, arising under cooling below $T_A$, is
commonly viewed as a result of superposition of relaxation processes
with an extended range of relaxation times. Thus, the additional
relaxation stretching detected using c(t) indicates an additional slow
relaxation process not captured by $F(Q_m,t)$.

\begin{figure} 
\includegraphics[width=8.1cm]{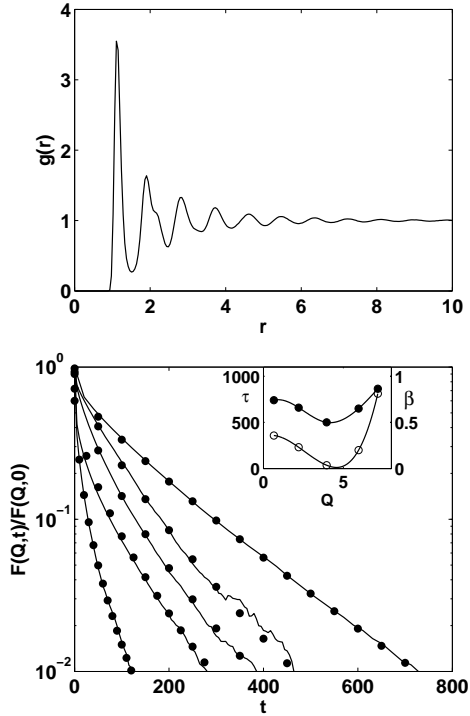}
\caption{ Top panel: radial distribution function for the simulated
  liquid at $T=0.7$. Bottom panel: $F(Q,t)$. From left to right: $Q=4,
  Q=6$,  $Q=2$,  $Q=0.6$, $Q=7.2$. Dots correspond to the fits of KWW. Inset
  shows the KWW parameters derived from the fitting. Open circles:
  $\tau$ (left-hand scale); dots: $\beta$ (right-hand scale).}
\label{fig4}
\end{figure}

We remind that a system with a finite configurational entropy
possesses a finite range of structural correlations. Therefore, the
ergodicity-restoring structural relaxation in a liquid is a strictly
local process. In the case of pair correlations, the correlation range
$r_c$ corresponds to vanishing radial distribution function $g(r)$.
Respectively, if relaxation of the pair correlations is considered in
terms of $F(Q,t)$, the lowest $Q$ relevant for the structural
relaxation can be estimated as $Q_{min}=2 \pi/r_c$. $Q$-dependent
relaxation time of $F(Q,t)$ can be estimated from the structure factor
$S(Q)$\cite{degennes} as $\tau(Q)=S(Q)/Q^2$. In a typical dense simple
liquid \cite{Hansen} $S(Q_m)/S(Q_{min}) > (Q_m/Q_{min})^2$, and,
therefore, $\tau(Q_m)$ is expected to exceed $\tau(Q_{min})$. In Fig.
4, we test this conjecture for the simulated liquid at $T=0.7$. The
figure shows that $g(r)$ apparently vanishes beyond $r_c=10$, from
which we estimate $Q_{min}\approx0.6$. Relaxation of $F(Q,t)$ is
analysed within the relevant range of $Q$ bounded by $Q_{min}$ and
$Q_m$. For each value of $Q$, the decay of $F(Q,t)$ is quantified by
the KWW fit; the $Q$ variations of the resulting KWW parameters are
shown in the inset. Evidently, $\tau(Q_m)$ is a pronounced absolute
maximum within the explored range of $Q$, whereas $\beta$ doesn't
change significantly. Therefore, structural relaxation in our liquid
at $T=0.7$, in the pair approximation, is entirely controlled by
$F(Q_m,t)$. This leads to the conclusion that the observed delay of
the total structural relaxation relative to the decay of $F(Q_m,t)$ is
due to the delay of the relaxation of higher-order correlations
relative to that of pair correlations.

The main peak of $S(Q)$ is linked to the local order (cages of nearest
neighbours). Therefore, one reason for the decay of $F(Q_m,t)$ can be
a comprehensive dissipation of the local order due to uncorrelated
particle motions. But it can also be caused by collective particle
movements preserving the higher-order structural correlations. This
will delay the total structural relaxation relative to the decay of
$F(Q_m,t)$ as we observe for $T=0.7$. The increase in the life-time of
these structural correlations can be attributed to a rapid buildup of
locally preferred structure that is expected upon cooling the liquid
towards $T_c$. This transformation in the relaxation dynamics can be
interpreted as a crossover to the regime of compact CRR as conjectured
in Ref.~\cite{Wolynes}.

The locally preferred structure in this system is icosahedral. Like an
earlier studied liquid with a similar structure \cite{dzugutov,
glotzer}, it demonstrates a strong tendency for a low-dimensional
icosahedral aggregation \cite{Z2} growing towards percolation as $T
\rightarrow T_c$. Relative mobility of these structural elements can
be conjectured as a conceivable reason for the decoupling between
$c(t)$ and $F(Q_m,t)$. As a corroborating observation, the anomalous
decay of $F(Q_m,t)$ reported here can be compared with a related
$Q$-dependent anomaly in the non-ergodicity parameter in a supercooled
low-density liquid approaching gelation transition \cite{sciortino}
caused by a percolating cluster network. Compact 3D CRR were also found
below $T_c$ in the binary Lennard-Jones (BLJ) system \cite{vollmayr}
and NiZr\cite{teichler}, and bond-preserving movements of structural
domains was observed at a saddle point crossing of the BLJ system
\cite{wales}. On the other hand, a study of the frequency-dependent
specific heat in silica \cite{Scheidler}, also based on the analysis
of the kinetic-energy fluctuations, showed no discrepancy between the
relaxation of these fluctuations and the decay of the {\it self part}
of $F(Q,t)$.

We note that, because of the particles' indistinguishability, the
structural correlations can survive particles exchange induced by
vacancy hopping. Therefore, the slow energy fluctuations we report
here cannot be necessarily associated with breaking bonds connecting
the nearest neighbours. The vacancy-assisted particle hopping, common
in supercooled liquids, significantly complicates the problem of
identifying the groups of particles involved in the conjectured compact
CRR dynamics.

One of the consequences of compact CRR  concerns the relaxation
of shear stress. The liquid can possibly accommodate to the shear
strain, and relax the induced stress by a mutual rearrangement of
slowly relaxing clusters \cite{stillinger} before their internal
structure has dissipated. Thus, shear stress relaxation in a
supercooled liquid, like the dissipation of the local density
fluctuations as measured by $F(Q_m,t)$, doesn't necessarily imply
structural relaxation.

In summary, we found a novel aspect of the supercooled liquid
dynamics manifested by the advanced decay of $F(Q_m,t)$ relative to
the actual structural relaxation. The latter has been assessed from
the decay of the time correlations of the system's total kinetic
energy fluctuations. The observed effect can be interpreted as being
caused by the cooperative movements of particles constrained by
slowly-decaying many-body structural correlations within compact CRR
which appear as the liquid is cooled close to $ T_c$. The observed
transformation in the relaxation mechanism can be regarded as a result
of the metabasin topography of the PEL of our model.

The authors thank Prof. D. J. Wales for 
very useful discussions. The support from the Centre for Parallel
Computers (PDC) is gratefully acknowledged.

\end{document}